\documentclass[review]{elsarticle}

\usepackage{lineno,hyperref}

\journal{}

\usepackage{numcompress}\biboptions{authoryear}

\begin{document}

\begin{frontmatter}

\title{Stationarity analysis of the stock market data and its application to mechanical trading}

\author[label1]{Kazuki Kanehira}

\author[label2]{Norikazu Todoroki}
\ead{todoroki.norikazu@p.chibakoudai.jp}

\address[label1]{Department of Computer Science, Chiba Institute of Technology 2-17-1 Tsudanuma, Narashino, 275-0016 Chiba, Japan}
\address[label2]{Physics Division, Chiba Institute of Technology 2-1-1 Shibazono, Narashino, 275-0023 Chiba, Japan}

\begin{abstract}
This study proposes a scheme for stationarity analysis of stock price fluctuations based on KM$_2$O-Langevin theory. Using this scheme, we classify the time-series data of stock price fluctuations into three periods: stationary, non-stationary, and intermediate. We then suggest an example of a low-risk stock trading strategy to demonstrate the usefulness of our scheme by using actual stock index data. Our strategy uses a trend-based indicator, moving averages, for stationary periods and an oscillator-based indicator, psychological lines, for non-stationary periods to make trading decisions. Finally, we confirm that our strategy is a safe trading strategy with small maximum drawdown by back testing on the Nikkei Stock Average. Our study, the first to apply the stationarity analysis of KM$_2$O-Langevin theory to actual mechanical trading, opens up new avenues for stock price prediction.
\end{abstract}


\end{frontmatter}


\section{Introduction}
The stock market is believed to be efficient at reflecting information by allowing it to spread very quickly and be mirrored in the stock prices. This theory is called the efficient market hypothesis. Because of this hypothesis, stock price fluctuation is a random walk process, and it is considered impossible to predict stock prices based on historical stock prices (\cite{Fama1970}).

Notwithstanding this belief, many studies on stock price forecasting have been conducted using time series analysis and multiple regression models. In addition, machine learning techniques such as neural networks (NNs) and genetic algorithms (GAs) have also been applied in this field. For example, \cite{Atsalakis2009,Atsalakis2013} reviewed several papers on stock price forecasting. Researchers have also attempted to find strategies to overcome the efficient market hypothesis. \cite{Booth2014} point out that although many of these studies on stock price forecasting show that we can make more profits than the benchmark, most schemes exhibit large drawdowns or a vast number of trades, resulting in very high transaction costs.
 
No deterministic schemes have been found so far because the time series of stock prices often behave as a random walk process, as predicted by the efficient market hypothesis. For example, a strategy is optimised on a learning interval that includes a theoretically unpredictable period stock price using a machine learning technique. In this case, the obtained strategy is no longer able to obtain a high-profit strategy in the test data. 

However, when we look at historical charts of stock prices, we see periods in which stock prices behave differently from random walks. Examples include the 2008 financial crisis and the 2020 stock market crash. Furthermore, it is well known that the distribution of the continuously compounded return (CCR) yield is not a Gaussian distribution obtained from a random walk process but rather a fat-tail distribution (\cite{Kariya1993}). 
 
Even models that consider deviations from the Gaussian changes in volatility have been proposed. For example, the ARCH model proposed by Engle, and the improved version of the ARCH model proposed by Bollerslev, the GARCH model, are well known and widely used (\cite{Engle1982,Bollerslev1986}).

However, some studies consider this deviation from the Gaussian distribution as non-stationary of time series data. One of the methods to examine the stationarity of time-series data is the Test(S) in KM$_2$O-Langevin theory (\cite{Okabe1999,Okabe2000,OkabeandNakano1991,OkabeandInoue1994,OkabeandYamane1998}). \cite{Sakai2005} applied the Test(S) in KM$_2$O-Langevin theory to historical data of stock prices. Moreover, \cite{Suzuki2006} analysed the stationarity of historical stock price data using a method called Test(ABN), an application of Test(S). As a result, they detected the beginning of the period when the stock price deviates from the random walk, which causes large fluctuations, by Test(ABN). However, it is crucial to know early the start and the end of this period to find a profitable strategy. When we know in advance when a stock price deviates from the random walk process, we expect to optimise the high-profit strategy effectively. Therefore, we improve Test(ABN) and propose a method of stationarity test that can determine the beginning and the ending of the period earlier. In addition, we show an example of applying the proposed method to a trading strategy. Finally, we test this strategy using historical data from the Nikkei Stock Average (Nikkei Stock Average is owned by Nikkei Inc.; the data were obtained from https://indexes.nikkei.co.jp). The present stationarity analysis and trading strategy have considerable forecasting power across the Nikkei stock average. To the best of our knowledge, this is the first study that applies stationarity analysis to the KM$_2$O-Langevin theory to an actual mechanical trading system.

The remainder of this paper is organised as follows. In Section 2, we describe the proposed stationarity analysis. In Section 3, we propose a method for a trading strategy using stationarity analysis. Finally, section 4 presents the discussion and conclusions.

\section{Stationarity analysis}
\begin{figure}[tb]
\centering

\includegraphics[scale=0.75]{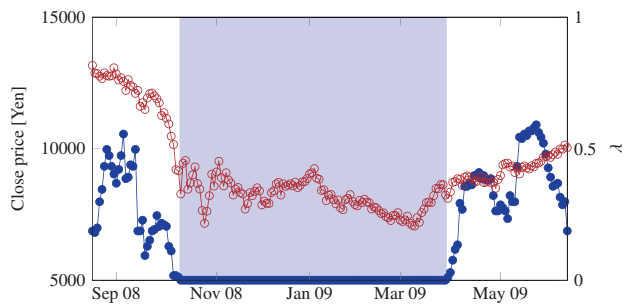}

{\footnotesize (a)}

\vspace{1em}

\includegraphics[scale=0.75]{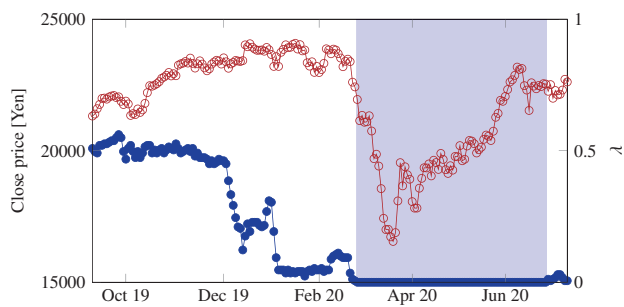}

{\footnotesize (b)}

\caption{Close price of the Nikkei Stock Average at the 2008 financial crisis (a) and the 2020 stock market crash (b), and the values of stationary parameters $\lambda$ at that time. The red open circle represents the price and the blue solid circle represents $\lambda$, respectively. The periods judged as non-stationarity are shaded in blue. \label{original_KM2O_2008and2020}}
\end{figure}

This section describes the stationarity analysis. In the Black-Scholes model, CCR has a Gaussian distribution; therefore, in this study, the time-series data of daily stock prices are converted to the CCR as follows: 
\begin{eqnarray}
x(n)={\rm log} {\rm (Close\,\, price)}(n)-{\rm log} {\rm (Close\,\, price)}(n-1),
\end{eqnarray}
where ${\rm (Close\,\, price)(}n{\rm )}$ is the closing price on nth day. In practice, the CCR follows a fat tail distribution rather than a Gaussian distribution because it fluctuates significantly. Our objective is to classify stock price fluctuations from the viewpoint of stationarity and to develop trading strategies that eliminate dangerous periods.
To examine the stationarity of the CCR on $i$th day, we first cut out $N+1$ data from $(i-N)$th day to $i$th day of the CCR: $x^{(i)}=(x^{(i)}(n);0\le n\le N)$ with $x^{(i)}(n)=x(i+n-N)$. We set $N=100$. We then normalise the time series $x^{(i)}(n)$ by 
\begin{eqnarray}
\tilde{x}(n)=\frac{x^{(i)}(n)-\mu (x^{(i)})}{\sqrt{v(x^{(i)})}},
\end{eqnarray}
where $\mu (x^{i})$ and $v(x^{i})$ are the sample mean and sample variance of the time series $x^{(i)}(n)$, respectively, defined as
\begin{eqnarray}
\mu  (x^{(i)})&\displaystyle =&\frac{1}{N+1}\sum_{n=0}^{N} x^{(i)}(n), \\
v (x^{(i)})&\displaystyle =&\frac{1}{N+1}\sum_{n=0}^{N}\left ( x^{(i)}(n)-\mu (x^{(i)})\right )^2.
\end{eqnarray}
Next, we perform 19 non-linear transformations to determine the properties behind time-series data:
\begin{eqnarray}
\left \{
\begin{array}{ll}
x_0&=(\tilde{x}(n);0\le n\le N),\\
x_1&=(\tilde{x}(n)^2;0\le n\le N),\\
x_2&=(\tilde{x}(n)^3;0\le n\le N),\\
x_3&=(\tilde{x}(n)\tilde{x}(n-1);1\le n\le N),\\
x_4&=(\tilde{x}(n)^4;0\le n\le N),\\
x_5&=(\tilde{x}(n)^2\tilde{x}(n-1);1\le n\le N),\\
x_6&=(\tilde{x}(n)\tilde{x}(n-2);2\le n\le N),\\
x_7&=(\tilde{x}(n)^5;0\le n\le N),\\
x_8&=(\tilde{x}(n)^3\tilde{x}(n-1);1\le n\le N),\\
x_9&=(\tilde{x}(n)^2\tilde{x}(n-2);2\le n\le N),\\
x_{10}&=(\tilde{x}(n)\tilde{x}(n-1)^2;1\le n\le N),\\
x_{11}&=(\tilde{x}(n)\tilde{x}(n-3);3\le n\le N),\\
x_{12}&=(\tilde{x}(n)^6;0\le n\le N),\\
x_{13}&=(\tilde{x}(n)^4\tilde{x}(n-1);1\le n\le N),\\
x_{14}&=(\tilde{x}(n)^3\tilde{x}(n-2);2\le n\le N),\\
x_{15}&=(\tilde{x}(n)^2\tilde{x}(n-1)^2;1\le n\le N),\\
x_{16}&=(\tilde{x}(n)^2\tilde{x}(n-3);3\le n\le N),\\
x_{17}&=(\tilde{x}(n)\tilde{x}(n-1)\tilde{x}(n-2);2\le n\le N),\\
x_{18}&=(\tilde{x}(n)\tilde{x}(n-4);4\le n\le N).
\end{array}
\right .
\end{eqnarray}
We denote by $\sigma_i$ the left-hand side of the time domain of the time series $x_i(n)$. We select two of the 19 time-series data obtained by non-linear transformation and construct two-dimensional time-series data as follows:
\begin{eqnarray}
x_{(i,j)}&=&\left ( x_i(n), x_j(n);\sigma_{(i,j)}<n<N\right ), (0\le i<j\le 18),\\
\sigma_{(i,j)}&=&{\rm max}(\sigma_i,\sigma_j).
\end{eqnarray}
Then, we perform the stationarity analysis, Test(S) on KM$_2$O-Langevin theory on the two-dimensional time-series data $x_{(i,j)}$. We briefly summarise the KM$_2$O-Langevin theory and Test(S) in \ref{appendixa}. There were 171 independent two-dimensional time series. In Test(ABN), we define the following stationarity parameter:
\begin{eqnarray}
\lambda (i)=\frac{\rm Number\, of\, time\, series\, pairs\, which\, passed\, criterion\, Test(S)}{\rm Total\, number\, of\, time\, series\, pairs : 171}.
\end{eqnarray}
The range of values for $\lambda$ ranged from 0 to 1.0. If the $\lambda$ is finite on $i$th day and zero on $(i+1)$th day, we say that $(i+1)$th day is the beginning of the non-stationary period (\cite{Nakamula2007, Suzuki2006}).
 Figure \ref{original_KM2O_2008and2020} shows the close price of the Nikkei Stock Average at the time of the financial crisis of 2008 and the stock market plunge of 2020, and the values of stationarity parameters  at that time. We shade blue the periods determined to be non-stationary. 

One problem of Test(ABN) is that Test(ABN) judges the interval where the stock price falls significantly just before the non-stationary period in the 2008 stock market collapse as stationary. Traders are likely to incur substantial losses during this  interval. Therefore, a scheme to determine this interval of large decline in advance is required.

Another problem is that Test(ABN) can determine the beginning date of the non-stationary period, but not the end date exactly. For the calculation of the  on $i$th day, we used the data from $(i-N+1)$th to $i$th day. For the calculation of the  on $(i+N-1)$th day, we used the data from $i$th day to $(i+N-1)$th day. Thus, if there is significant change in stock price only on the $i$th day, it impacts on determining stationarity until $(i+N-1)$th day. This means that the estimation of the end date of the non-stationary interval is shifted by at most $(N-1)$ days. 

The conventional Test(S) criterion for the stationarity of time-series data is given in (\ref{criteriaS}). Our objective is not to strictly determine the stationarity of stock price data. Therefore, we modified this criterion to solve the following problems.
\begin{eqnarray}
\label{criteriaS2}
{\rm Test(S)'}=\left \{
\begin{array}{ll}
{\rm The\,\, rate\,\, that\,\, criterion\,\, }({\rm M})_{s}& \hspace{-0.6em}{\rm is\,\, passed\,\, exceeds\,\, (80\times \alpha)~\%.}\\
{\rm The\,\, rate\,\, that\,\, criterion\,\, }({\rm V})_{s}
&\hspace{-0.6em} {\rm is\,\, passed\,\, exceeds\,\, (70\times \alpha)~\%.}\\
{\rm The\,\, rate\,\, that\,\, criterion\,\, }({\rm O})_{s}
&\hspace{-0.6em} {\rm is\,\, passed\,\, exceeds\,\, (80\times \alpha)~\%.}
\end{array}
\right .
\end{eqnarray}
where $\alpha$ is a constant that takes a value between zero and one. When $\alpha=1$, criterion (\ref{criteriaS2}) is the conventional Test(S). 

\begin{figure}[htbp]
\centering

\includegraphics[scale=1]{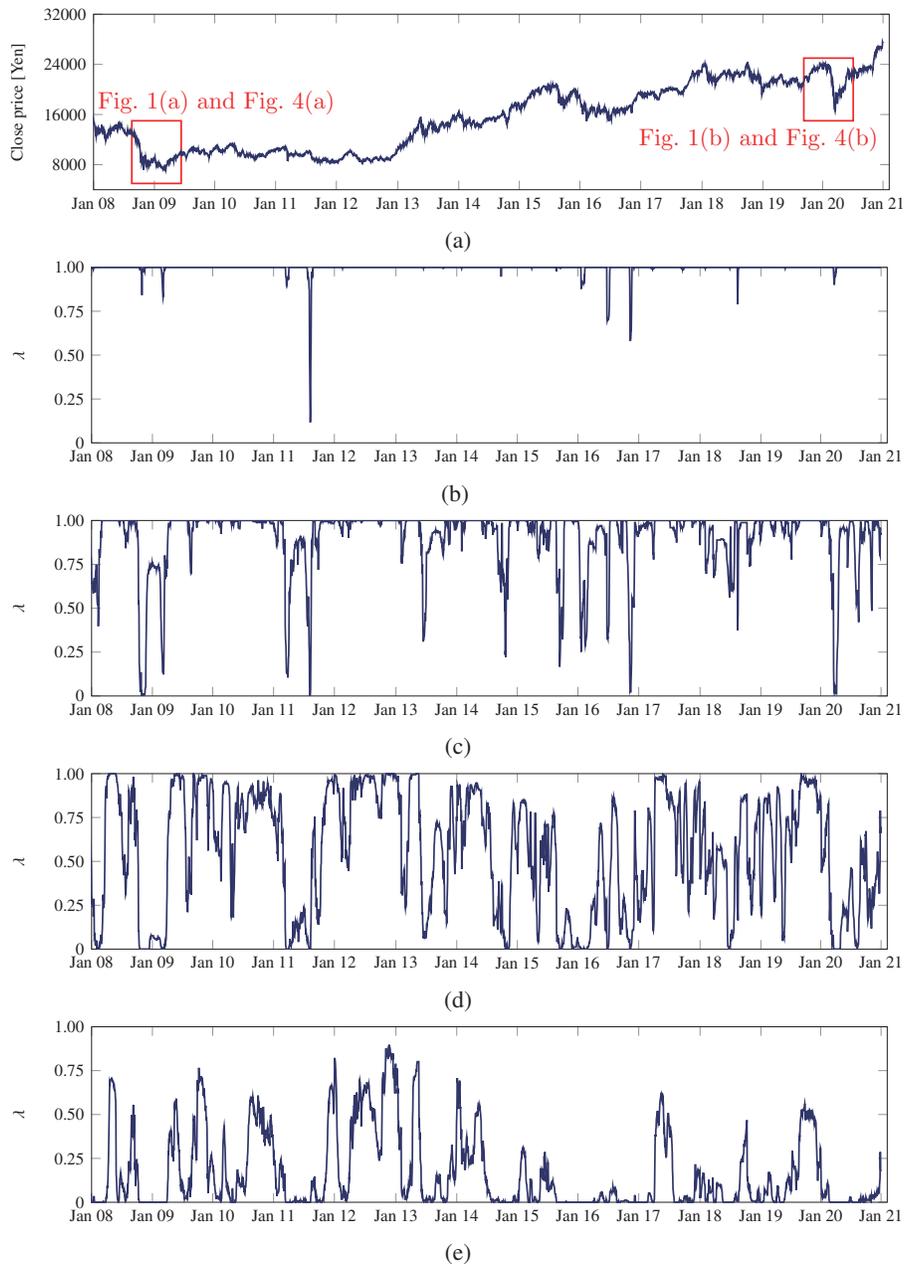}

\caption{Close price of the Nikkei Stock Average and its stationarity parameter. (a) is evaluation of close price. (b), (c), (d), and (e) are the stationarity parameter when $\alpha=0.25, 0.5, 0.75$, and, $1.0$, respectively. $\alpha=1.0$ is the conventional Test(S). In (a), the part corresponding to the enlarged area in Fig. \ref{original_KM2O_2008and2020} and Fig. \ref{proposed_KM2O_2008and2020} is represented by a red square.\label{lambda_parameter} }
\end{figure}

We show the changes over time of the stationarity parameters $\lambda$ for the Nikkei Stock Average with various $\alpha$ in Figure \ref{lambda_parameter}. The smaller the value of  $\alpha$, the closer the value of $\lambda$ is to 1.0 in the stationary period. When the $\alpha$ value is close to 1, $\alpha$ stretches the $\lambda$ value in the vertical direction. Contrarily, when we further reduce the $\alpha$ value, the $\lambda$ value approaches one even in the non-stationary period. Therefore, it is necessary to set an appropriate $\alpha$ value to determine the non-stationary period and the fall in stock price just before the non-stationary period. For our purposes, the value of $\alpha$, where $\lambda$ touches 1 only in the stationary period, is appropriate.

\begin{figure}[tb]
\centering

\includegraphics[scale=0.75]{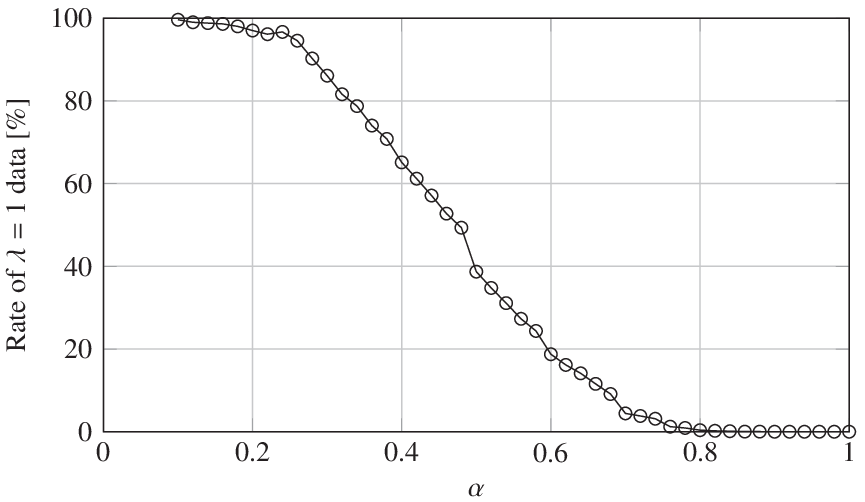}

\caption{Rate of data with $\lambda=1$ in the Nikkei Stock Average\label{rateoflambdaeq1}}
\end{figure}

Figure \ref{rateoflambdaeq1} shows the rate of days when $\lambda(i)=1.0$ for different $\alpha$ values in the Nikkei Stock Average. Looking at this figure, it seems that $\alpha=0.3\sim 0.6$ is appropriate. In the following, we set $\alpha= 0.5$. Figure \ref{lambda_parameter}(c) shows the stationarity parameter with $\alpha=0.5$ of the Nikkei stock average, where $\alpha = 0.5$ seems to be a good value for the whole period, even including different stock price behaviours. Therefore, we expect that this value will not vary significantly among stocks.

After determining the value of $\alpha$, we set the two thresholds of $\lambda$ to distinguish between the three periods with different behaviours. The first is $\lambda_1^{*}$, which is the threshold for whether $\lambda$ moves away from 1. We set $\lambda_1^{*}$ to remove the small fluctuations. Therefore, we need to set the value of $\lambda_1^{*}$ to a value slightly less than 1. We refer to the period of $\lambda\ge \lambda_1^{*}$ stationary period. Here, we set $\lambda_1^{*}= 165.5/171\sim 0.968$. Another threshold, $\lambda_2^{*}$, is the threshold for determining the non-stationary periods. We refer to the period $\lambda<\lambda_2^{*}$ a non-stationary, and the period  $\lambda_1^{*}>\lambda\ge \lambda_2^{*}$ an intermediate, that is, between non-stationarity and stationarity. Here, we set $\lambda_2^{*}= 100.5/171\sim 0.588$. If the values of $\lambda_1^{*}$ and $\lambda_2^{*}$ are set in this way, the Nikkei average from 2008 to 2020 will have 58.0\% of days in stationary, 34.6\% in intermediate, and 7.4\% in non-stationary. The kurtosis of CCR for all data was 7.98, and the kurtosis of CCR for stationary data was only 0.714. Because the kurtosis due to the Gaussian distribution is 0, we can confirm that the stationary data are sufficiently close to the Gaussian distribution.

\begin{figure}[tb]
\centering

\includegraphics[scale=0.75]{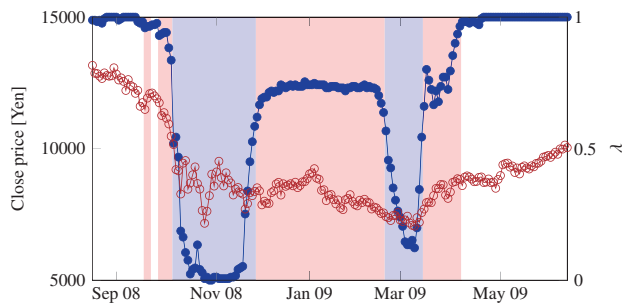}

{\footnotesize (a)}

\vspace{1em}

\includegraphics[scale=0.75]{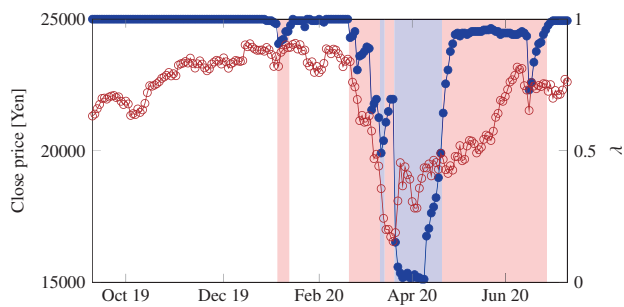}

{\footnotesize (b)}

\caption{Close price of Nikkei Stock Average at the 2008 financial crisis and the 2020 stock market crash, and the values of stationarity parameters $\lambda$ when $\alpha=0.5$ at that time. The red open circle represents the price and the blue solid circle represents $\lambda$, respectively. The periods judged to be non-stationary are shaded in blue, and the intermediate periods are shaded in red.\label{proposed_KM2O_2008and2020}}
\end{figure}

Figure \ref{proposed_KM2O_2008and2020} shows the Nikkei Stock Average during the financial crisis in 2008 and the stock market crash in 2020, and the values of the stationarity parameters $\lambda$ for $\alpha=0.5$ at those times. The periods judged to be non-stationary are shaded in blue, and the intermediate periods are shaded in red. Comparing Fig. \ref{original_KM2O_2008and2020} and Fig. \ref{proposed_KM2O_2008and2020}, it can be confirmed that, compared to the conventional Test(ABN), the proposed method determines the beginning of the intermediate period earlier than the beginning of the non-stationary period. Additionally, compared to the conventional Test(ABN), the proposed method judges the end of the non-stationary period earlier than the end of the non-stationary period.

\section{Application to mechanical trading}
We now present an example of the application of stationarity analysis to a real stock trading strategy. From the value of $\lambda$, we cannot determine whether the stock price will rise or fall. Therefore, we consider strategies that combine the technical indicators. Various indicators have been developed to date. We used the moving average and the psychological lines. The moving average is the most basic technical indicator and helps smooth out price behaviour by filtering out the noise from random price fluctuations. The $N_{\rm MA}$ days moving average on $i$th day $MA(N_{\rm MA},i)$ is as follows:
\begin{eqnarray}
MA(N_{\rm MA},i)=\frac{1}{N_{\rm MA}}\sum_{k=0}^{N_{\rm MA}-1}({\rm Close\, price})(i-k).
\end{eqnarray}
The psychological line as an indicator is the percentage of days that the stock price increases during the period under consideration. The $N_{\rm psy}$ days psychological line at $i$th day $Psy(N_{\rm psy},i)$ is calculated as follows:
\begin{eqnarray}
&&\hspace{-3em}Psy(N_{\rm psy},i)\nonumber \\
&&\hspace{-3em}=\frac{{\rm Number\, of\, times\, the\,  price\,  increased\,  from\,  }(i-N_{\rm psy}-1){\rm th\,  day\,  to\,  }i{\rm th\,  day}}{N_{\rm psy}}. 
\end{eqnarray}
In other words, if the psychological line is greater than 0.5, it indicates overbought, and if the psychological line is less than 0.5, it indicates oversold.
We use these indicators to develop a strategy consisting of the following three rules.

\begin{description}
\item{\bf Rule 1}: This rule is for the stationary periods. Price movement is a random walk-like fluctuation around the trend line in the stationary period. Because random walk-like fluctuations are unpredictable, we decide whether to buy or sell a position depending on the trend. We measure the trend based on the slope of the moving average, $MA(N_{\rm MA},i)-MA(N_{\rm MA},i-N_{\rm MA})$. That is, we decide with $MA(N_{\rm MA},i)-MA(N_{\rm MA},i-N_{\rm MA})\ge 0$ indicates a buy position, and with $MA(N_{\rm MA},i)-MA(N_{\rm MA},i-N_{\rm MA})<0$ indicates a sell position for the stationary periods. Here we set $N_{\rm MA}=5, 6, \cdots, 30$.

\item{\bf Rule 2}: This rule is for non-stationary periods, when there is a possibility that the fluctuations have some characteristics that are different from the random walk. Therefore, we try to obtain profit using psychological lines. In other words, we decide $Psy(N_{\rm psy},i)\le (N_{\rm psy}-1)/(2N_{\rm psy})$  indicates the buy position, and $Psy(N_{\rm psy},i)\ge (N_{\rm psy}+1)/(2N_{\rm psy})$ indicates the sell position for the non-stationary periods. Here, we set $N_{\rm psy}=3,5,7, 9$, and $11$.

\item{\bf Rule 3}: This rule is for the intermediate periods, where the fluctuation may be random walk-like or non-random walk-like. We may be able to make much profit by using an appropriate technical indicator during in this period. However, if we fit the price movements of different behaviours with a single rule, there will be unreasonableness somewhere, which may lead to an increase in the maximum drawdown. Therefore, we decide to be in no position for the intermediate periods.
\end{description}

We used these rules to determine our positions and then executed the trades the day after the positions were changed. Note that we only use the information from the previous day's closing price for that day's trade. For simplicity, we only trade one share and assume that the initial capital is infinite and there is no stock trading fee. 

\begin{figure}[tb]
\centering

\includegraphics[scale=1]{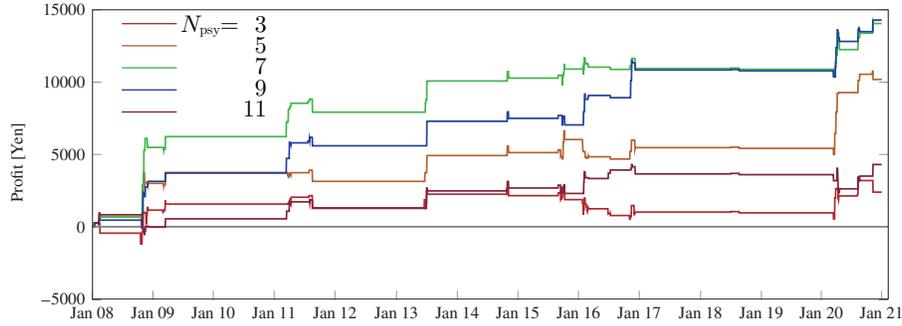}

{\footnotesize (a)}

\vspace{1em}

\includegraphics[scale=1]{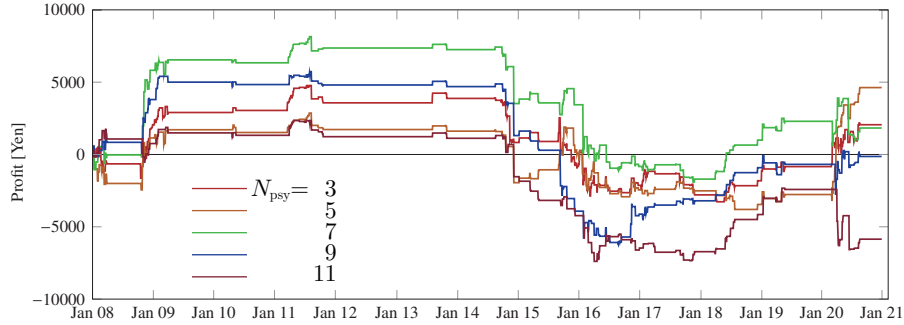}

{\footnotesize (b)}

\caption{Profit curves of the Rule 2-only strategy on the Nikkei Stock Average. (a) shows the results of trading only in the non-stationarity period judged by the proposed method. (b) shows the results of trading only in the non-stationarity period judged by the conventional Test(ABN) and traded only during non-periodic periods. Red, orange, green, blue, and purple are using the psychological lines with $N_{\rm psy}=3, 5, 7, 9$, and $11$, respectively. \label{onlyrule2}}
\end{figure}

We demonstrate our proposed strategy using the Nikkei Stock Average. First, to assess the usefulness of Rule 2, we consider only Rule 2 for the non-stationary period and execute the trade setting out that the other periods are in no position. Figure \ref{onlyrule2}(a) shows the time evolution of the fixed profit when only Rule 2 is considered. Figure \ref{onlyrule2}(b) shows, for comparison, the results of trading using Rule 2 during the non-stationary period judged by the conventional  Test(ABN). In Fig. \ref{onlyrule2}(a), we can see that profits are generated with $N_{\rm psy}=7$ and 9 in almost all periods. This result suggests that we may predict short-time fluctuations in the non-stationary period. Furthermore, the best performance is obtained at $N_{\rm psy}=9$, suggesting that the fluctuations in the non-stationary periods are fluctuations with a characteristic length scale of approximately 9 days. In Fig. \ref{onlyrule2}(b), we can see that the loss occurs in the latter half of the period when the conventional Test(ABN) is used. In the conventional Test(ABN), the period considered non-stationarity is wider than that of the proposed method, so the unpredictable period is probably included in the non-stationary periods.

\begin{figure}[tb]
\centering

\includegraphics[scale=1]{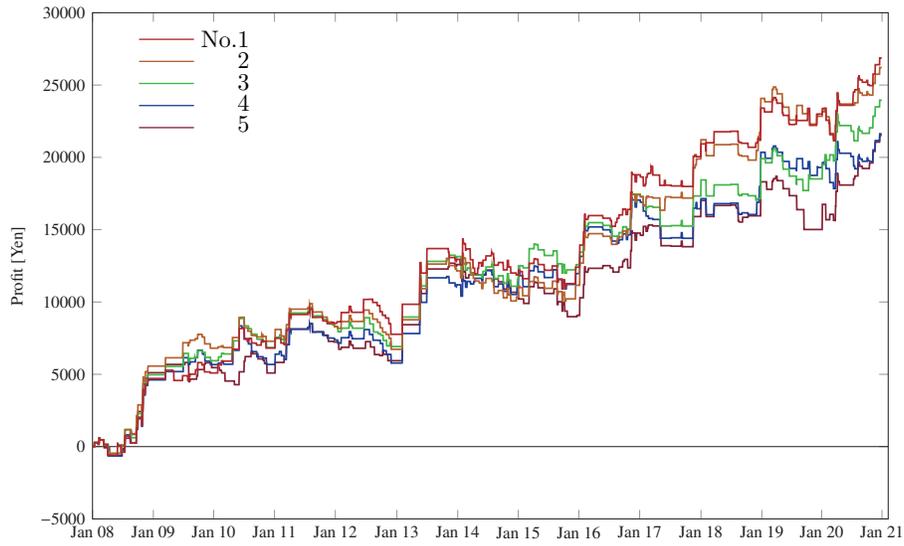}

{\footnotesize (a)}

\vspace{1em}

\includegraphics[scale=1]{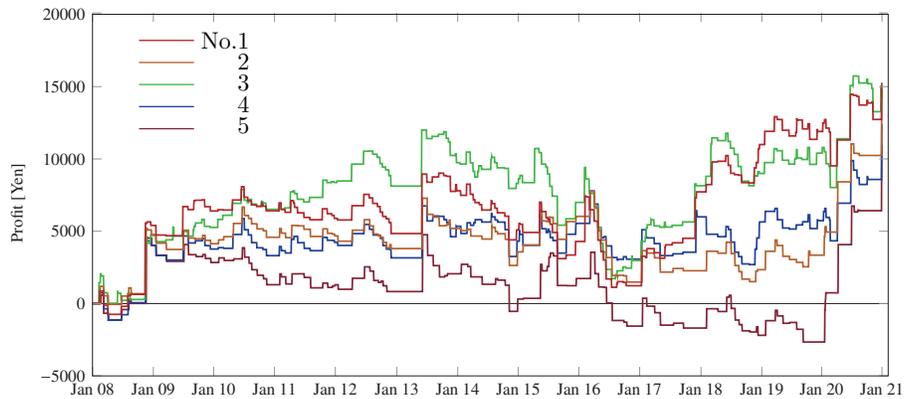}

{\footnotesize (b)}

\caption{Profit curves on the Nikkei Stock Average. (a) shows the results of the proposed strategy. (b) shows the results of all periods are traded according the moving average trend. Red, orange, green, blue, and purple are No. 1, 2, 3, 4, and 5 in Table 1, respectively. \label{profitcureve}}
\end{figure}

Finally, we fixed $N_{\rm psy}=9$, changed $N_{\rm MA}$, and tested the performance of the strategy by considering all the rules. We then chose the five values of $N_{\rm MA}$ with the high-profit strategies. The time evolution of the fixed profit of these strategies is shown in Figure \ref{profitcureve}(a). In addition, Table \ref{beststrategy}(a) shows the number of trades ($N_{\rm trade}$), profit, maximum drawdown (MDD), and profit factor (PF) of these strategies. Figure \ref{profitcureve}(b) and Table \ref{profitcureve}(b) show, for comparison, traded in all periods according to the trend determined from the moving average, as in Rule 1.  The MDD of the strategies in Table \ref{beststrategy}(b) is large. This is because trading with a single rule on periods of different fluctuation has become unreasonable. On the other hand, the results in Fig. \ref{profitcureve}(a) and Table \ref{beststrategy}(a) show that when trading with the proposed strategy, the profit steadily increases over the entire period  with a small MDD.

\begin{table}
\centering 
\caption{The number of trades ($N_{\rm trade}$), profit, maximum drawdown (MDD), and profit factor (PF) of these strategies. (a) shows the results of proposed strategy. (b) shows the results of all periods are traded according to the trend determined from the moving average.\label{beststrategy}}

(a) 

\begin{tabular}{rrrrrr}
\hline 
No. & $N_{\rm MA}$ & $N_{\rm trade}$ & Profit~[Yen] & PF & MDD[Yen] \\ \hline \hline 
1 & 10 &  233 &  26873.96 & 1.73 & 3475.59 \\
2 & 11 & 221 & 26216.10 & 1.72  & 3626.87 \\
3 & 13 & 205 & 23954.48 & 1.68 & 2884.37 \\
4 & 12 & 218 & 21627.56 & 1.56 & 2944.09 \\
5 & 17 & 187 & 21525.23 & 1.63 & 3697.84\\ \hline 
\end{tabular}
\vspace{1em}

(b) 

\begin{tabular}{rrrrrr}
\hline 
No. & $N_{\rm MA}$ & $N_{\rm trade}$ & Profit~[Yen] & PF & MDD[Yen] \\ \hline \hline 
1 & 11 & 160 &  15268.47 & 1.38 & 7894.45 \\
2 & 13 & 126 & 14804.25 & 1.44  & 6259.91 \\
3 & 10 & 181 & 14232.40 & 1.31 & 10274.95 \\
4 & 12 & 146 & 12405.03 & 1.31 & 5093.65 \\
5 & 14 & 116 & 10986.04 & 1.32 & 7471.69 \\ \hline 
\end{tabular}

\end{table}

\clearpage 

\section{Discussion and Conclusion}
In this study, we proposed a practical scheme for a mechanical trading system by improving the Test(ABN) of the KM$_2$O-Langevin theory. Using the proposed scheme, we classified the time-series data of stock price fluctuations into three periods: stationary, non-stationary, and intermediate. As an application of our proposed scheme to mechanical trading, we devised a simple strategy that combines moving averages and psychological lines and demonstrated with the Nikkei Stock Average. The proposed strategy consists of a rule using the moving average as a trend index for the stationary period, and a rule using the psychological line as an oscillator index for the non-stationary period. This strategy allows safe trading with a small MDD.

Finally, we want to emphasise that since the non-stationary period obtained in the present scheme is not a random walk, there is a possibility that we can adequately predict the rise and fall of the stock price. Although this study used a strategy to detect overbuying and overselling during non-stationary periods using psychological lines, many technical indicators have been proposed to date for such detection. Using other technical indicators may more efficiently detect fluctuations during non-stationary periods. It is not clear which technical indicator is the best to use, but we can use machine learning techniques such as GA and NN to find the best technical indicator. When investigating which technical indicator should be used with machine learning techniques, it is to be remembered that efficient learning will be possible by limiting it to non-stationary data that are not a random walk. The application of the proposed scheme with machine learning is a topic for future research.

\appendix 
\section{KM$_2$O-Langevin Theory and Stability Analysis}
\label{appendixa}
We briefly review the KM$_2$O-Langevin theory introduced by Okabe (\cite{Okabe1999,Okabe2000,OkabeandNakano1991,OkabeandInoue1994,OkabeandYamane1998}).
We consider the $N+1$ real-valued two dimensional stochastic process ${X}=\!(X(n);0\le n\le N)$ which is defined on the probability space $(\Omega,{\cal B},P)$.
Here, $X(n)$ are two-components vectors, all of which are square-integrable. 
${X}=\!(X(n);0\le n\le N)$ is called weak stationary process if $X$ satisfies the following relation.
\begin{eqnarray}
E[X(n)]&=&\mu,\\
E[(X(n)-\mu)^t(X(n')-\mu)]&=&R(t-s),\quad  (0\le n,n'\le N)
\end{eqnarray}
where $E[\cdot]$ denotes the expected value in the probability space $(\Omega,{\cal B},P)$.

When the Teplitz matrix constructed from the covariance function of ${X}(t)$ is regular, we can derive the stochastic differencial equation that describes the time evolution of the non-degenerate stochastic process $X$ as follows:
\begin{eqnarray}
\label{KM$_2$O01}
\left \{
\begin{array}{ll}
{X}(n)&\displaystyle =-\sum_{k=0}^{n-1}\gamma_{+}(n,k){X}(k)+{\nu}_{+}(n),\\
{X}(N-n)&\displaystyle =-\sum_{k=0}^{n-1}\gamma_{-}(n,k){X}(N-k)+{\nu}_{-}(-n),
\end{array}
\right .
\end{eqnarray}
with ${\nu}_{+}(0)= {\nu}_{-}(0)= {X}(0)$. 

The first term of Eq.\ref{KM$_2$O01} is the dissipation part, and the second term inof Eq.\ref{KM$_2$O01} representsis the fluctuation part. $\gamma_\pm$ is a $2\times 2$ matrix determined uniquely from the non-degenerate stochastic process $X$. $\gamma_\pm$ is called the KM$_2$O-Langevin dissipation matrix. $\nu_{\pm}$ is a $d$-dimensional vector that corresponds to the external driving force.

$\nu_+$ satisfies the following relation.
\begin{eqnarray}
E[{\nu}_+(n)]&=&0,\\
E[{\nu}_+(n)^t,{\nu}_+(n')]&=&\delta_{n,n'}{V}(n), \quad (0\le n,n'\le N),
\end{eqnarray}
where $\delta_{n,n'}$ is the Kronecker delta. $V(n)$ is a $2\times 2$ matrix which is also determined uniquely from the non-degenerate stochastic process ${X}$, and is called the KM$_2$O-Langevin fluctuation matrix. The necessary and sufficient condition for $X$ to have weakly stationary property is the dissipation-dissipation theorem and fluctuation-dissipation theorem.
\begin{eqnarray}
\left \{
\begin{array}{ll}
\gamma_{\pm}(1,0)&=\delta_{\pm}(1),\\
\gamma_{\pm}(n,0)&=\delta_{\pm}(n+1),\\
\gamma_{\pm}(n,k)&=\gamma_{\pm}(n-1,k-1)+\delta_{\pm}(n)\gamma_{\mp}(n-1,n-k-1),
\end{array}
\right .
\end{eqnarray}
and
\begin{eqnarray}
\left \{
\begin{array}{ll}
V_{\pm}(0)&=R(0),\\
V_{\pm}(n+1)&=\left (1-\delta_{\pm} (n+1)\delta_{\mp} (n+1)\right )R(n),\quad (1\le n\le N),\\
\end{array}
\right .
\end{eqnarray}
where $\delta$ are obtained from the covariance function $R$ of the stochastic process $X$.
\begin{eqnarray}
\left \{
\begin{array}{ll}
\delta_{\pm}(0)&=-R(\pm 1){R(0)}^{-1},\\
\delta_{\pm}(n+1)&\displaystyle =-\left \{
R(\pm (n+1))+\sum_{k=0}^{n-1}\gamma_{\pm}(n,k)R(\pm (k+1))
\right \}{V_{\mp}(n)}^{-1}.
\end{array}
\right .
\end{eqnarray}
Using these relation, we can derive the basic characteristic quantities  $\{\delta \gamma V\}$ of the KM$_2$O-Langevin equation from the covariance function $R$.

Test(S) is a scheme to investigate the stationarity of a stochastic process by extending the theory described above to real data and considering real data as a realisation of the stochastic process $X$.

We consider the real-time series data $Z=(Z(n);0\le n\le N)$. $Z(n)=^{t}(Z_1(n),Z_{2},\cdots ,Z_{d})$ are $d$-component vector.

The sample mean $\mu^Z$, and the sample covariance function $R^Z(n)$ for $Z(n)$ are defined by the following equations:
\begin{eqnarray}
\mu^Z&=&\frac{1}{N+1}\sum_{n=0}^{N}Z(n),\\
v_{ij}^Z&=&\frac{1}{N+1}\sum_{n=0}^{N}Z_i(n)^t\! Z_j(n),\\
R_{ij}^Z(n)&=&\frac{1}{N+1}\sum_{n=0}^{N}\left (Z_i(n+T)-\mu^Z\right )^t\!\left (Z_j(n)-\mu^Z\right ),
\end{eqnarray}
with 
\begin{eqnarray}
R^Z(n)&=^t\!R^Z(-n) ,
\end{eqnarray}
with $0\le n\le N$. First, we normalise $Z(t)$ as follows.
\begin{eqnarray}
\tilde{Z}(n)=\left (
\begin{array}{ccc}
\sqrt{{\left (v_{11}^Z\right )}^{-1}} & & 0\\
 &\ddots  &   \\
0 &  & \sqrt{{\left (v_{dd}^Z\right )}^{-1}}
\end{array}
\right )
{\left (Z(n)-\mu^Z\right )}.
\end{eqnarray}
By calculating the covariance function $R^z$ of the normalised time-series data $z$, we calculated the fundamental characteristic quantities. The stationarity of a time series $z$ means that $z$ is realised as a local, weakly stationary stochastic process with its covariance matrix $R^z$.

Data pieces of length $M+1$ are taken from data pieces of size $N$ and the stationarity of each piece is tested using $R=(R_{jk}(n);0\le n\le N,1\le j,k\le 2)$ calculated from the entire interval. An interval is defined as stationary when the ratio of data pieces that pass the stationarity check to all the data pieces in the interval exceeds a certain threshold.

Because the length $N$ is finite, the covariance matrix function is reliable only within a limited range. From the rule of thumb of time series analysis, among the $N+1$ values of the sample covariance matrix function, the valid numbers are in the range $[2\sqrt{N+1}/d]$ to $[3\sqrt{N+1}/d]$, where $[A]$ denotes a rounding down of $A$. Therefore, we set
\begin{eqnarray}
M=\left [3\sqrt{N+1}/d\right ]-1
\end{eqnarray}
and consider about $N-M$ data pieces of length $M$ for a fixed number of $s=0,1,\cdots ,N,-M$.

\begin{eqnarray}
\left \{
\begin{array}{lll}
\tilde{Z}^{(0)} & =  \left (\tilde{Z}(0), \tilde{Z}(1),\cdots ,\tilde{Z}(M)\right ), \\
\tilde{Z}^{(1)} & = \left (\tilde{Z}(1), \tilde{Z}(2),\cdots ,\tilde{Z}(M+1)\right ), \\
       &\vdots \\ 
\tilde{Z}^{(s)} & =  \left (\tilde{Z}(i), \tilde{Z}(i+1),\cdots ,\tilde{Z}(M+i)\right ), \\
      &\vdots&   \\
\tilde{Z}^{(N-M)} & =  \left (\tilde{Z}(N-M), \tilde{Z}(1),\cdots ,\tilde{Z}(N)\right ), \\
\end{array}
\right  .
\end{eqnarray}
From the sample covariance matrix $R^{\tilde{Z}}$, we calculated the fundamental characteristic quantities of the sample KM$_2$O-Langevin matrix system.

Using these quantities, we extract the sample forward KM$_2$O-Langevin fluctuation time series as
\begin{eqnarray}
\label{KM$_2$Onu001}
\nu_+^{(s)}(n)\equiv \tilde{Z}^{(s)}(n)+\sum_{k=0}^{n-1}\gamma_+(n,k)\tilde{Z}^{(s)}(k),\quad (0\le n\le M),
\end{eqnarray}

By the lower triangular matrix $W(n)$ such that $V_+(R^{\tilde{Z}})(n)=W(n)^tW(n)$, we normalise $\nu_{+}^{(s)}(n)$ to 
$
\xi_{+}^{(s)}(n)=W(n)^{-1}\nu_{+}^{(s)}(n)
$. 
In the case of $d=2$,
$W_{ij}(n)$, which is the $(i,j)$($i,j=1,2$) component of $W(n)$, is expressed using $V_{ij}(n)$, which is the $(i,j)$ component of $V_+(n)$, as follows:
\begin{eqnarray}
W_{11}&=&\sqrt{V_{11}(n)},\\
W_{12}&=&0,\\
W_{21}&=&\frac{V_{12}(n)}{\sqrt{V_{11}(n)}},\\
W_{22}&=&\frac{\sqrt{V_{11}(n)V_{22}(n)-{V_{12}}(n)^2}}{\sqrt{V_{11}(n)}}.
\end{eqnarray}
All components of $\xi(n)$ are arranged in one line, and the one-dimensional time series $\xi^{(s)}= (\xi^{(s)}(n); 0\le n\le d(M+1)-1)$ is constructed as
\begin{eqnarray}
\xi^{(s)}=\left (
\xi_{1}^{(s)}(0),\xi_{2}^{(s)}(0),
\xi_{1}^{(s)}(1),\xi_{2}^{(s)}(1),
\cdots,
\xi_{1}^{(s)}(M),\xi_{2}^{(s)}(M)
\right ).
\end{eqnarray}
$Z$ is the realisation of a local weakly stationary process with $R^Z$ as its covariance function is equivalent to the fact that $\xi$ is a realisation of standardised white noise. Therefore, we only need to test whether $\xi$ is a realisation of standardised white-noise. 

To test the white-noise condition of $\xi(s)$, we checked the normality and orthogonality of $\xi^{(s)}(n)$. 
The sample mean $\mu^{\xi^{(s)}}$, sample
pseudo-variance $v^{\xi^{(s)}}$, and sample pseudo-covariance $R^{\xi^{(s)}}(n;m)$ $(0\le n \le L, 0\le m\le L-n)$ are calculated as follows:
\begin{eqnarray}
\mu^{\xi^{(s)}}&=&\frac{1}{d(M+1)}\sum_{k=0}^{d(M+1)-1}\xi^{(s)}(k),\\
v^{\xi^{(s)}}&=&\frac{1}{d(M+1)}\sum_{k=0}^{d(M+1)-1}\xi^{(s)}(k)^2,\\
R^{\xi^{(s)}}(n,m)&=&\frac{1}{d(M+1)}\sum_{k=m}^{d(M+1)-1-n}\xi^{(s)}(k)\xi^{(s)}(n+k),
\end{eqnarray}
where $0\le n\le L$, $0\le m\le L-n$, and $L$ is the number of reliable data in the sample pseudo-covariance function $R^{\xi^{(s)}}(n,m)$. 
The criteria of sample mean, sample pseudo-variance, and sample pseudo-covariance function for the white noise condition are given as follows:
\begin{description}
\item {\bf Mean}: Inequality $\sqrt{2(M+1)} |\mu^{\xi^{(s)}}|<1.96$ is satisfied.
\item {\bf Variance}: Inequality $|(v^{\xi^{(s)}}-1)^\sim|<2.2414$ is satisfied.
Here, $(v^{\xi^{(s)}}-1)^\sim$ is defined by
\begin{eqnarray}
(v^{\xi^{(s)}}-1)^\sim=
\frac{\displaystyle \sum_{k=0}^{d(M+1)-1}(\xi^{(s)}(k)^2-1)}{\displaystyle \sqrt{\sum_{k=0}^{d(M+1)-1}(\xi^{(s)}(k)^2-1)^2}}.
\end{eqnarray}
\item{\bf Orthogonality}: The rate which an inequalities 
\begin{eqnarray}
2(M+1)\left (\sqrt{L_{n,m}^{(1)}}-\sqrt{L_{n,m}^{(2)}}\right )^{-1}|R^{\xi^{(s)}}(n,m)|<1.96,
\end{eqnarray}
is satisfied exceeds more than 90\%.
\end{description}

Let $d(M-1)-1$ and $m$ divided by $2n$ be $q$ and $u$, respectively, and let $r$ and $t$ be the remainders, respectively:
\begin{eqnarray}
d(M+1)-1&=q(2n)+r, \\
m&=u(2n)+r.
\end{eqnarray}
We define
\begin{eqnarray}
L_{n,m}^{(1)}&=&
\left \{
\begin{array}{ll}
n(q+u)-m, & \quad \quad \quad \quad(0\le t\le  n-1), \\
n(q-u-1), & \quad \quad \quad \quad(n\le t\le 2n-1).
\end{array}
\right .\\
L_{n,m}^{(2)}&=&
\left \{
\begin{array}{ll}
n(q-u-1)+r+1, & \quad (0\le t \le n-1), \\
n(q+u)+r+1-m, & \quad (n\le t\le 2n-1),
\end{array}
\right .
\end{eqnarray}
for $0\le r\le n$, and
\begin{eqnarray}
L_{n,m}^{(1)}&=&
\left \{
\begin{array}{ll}
n(q+u-1)+r+1-m, & \quad (0\le t \le n-1), \\
n(q-u-2)+r+1, & \quad (n\le t\le 2n-1),
\end{array}
\right .\\
L_{n,m}^{(2)}&=&
\left \{
\begin{array}{ll}
n(q-u), & \quad \quad \quad \quad (0\le t \le n-1) ,\\
n(q+u+1)-m, & \quad \quad \quad \quad(n\le t\le 2n-1),
\end{array}
\right .
\end{eqnarray}
for $n+1\le r\le 2n-1$.

Test(S) has been proposed as a criterion for determining whether data $Z$ can be regarded as a realisation value of a weakly stationary process.
\begin{eqnarray}
\label{criteriaS}
{\rm Test(S)}=\left \{
\begin{array}{ll}
{\rm The\,\, rate\,\, that\,\, criterion\,\, }({\rm M})_{s}& \hspace{-0.6em}{\rm is\,\, passed\,\, exceeds\,\, 80~\%.}\\
{\rm The\,\, rate\,\, that\,\, criterion\,\, }({\rm V})_{s}
&\hspace{-0.6em} {\rm is\,\, passed\,\, exceeds\,\, 70~\%.}\\
{\rm The\,\, rate\,\, that\,\, criterion\,\, }({\rm O})_{s}
&\hspace{-0.6em} {\rm is\,\, passed\,\, exceeds\,\, 80~\%.}
\end{array}
\right .
\end{eqnarray}
If Test(S) holds, we can regard time series $Z$ as the realisation value of the weakly stationary process. However, if Test(S) does not hold, we cannot regard the time series $Z$ as the realisation value of the weakly stationary process.


\begin{thebibliography}{15}
\expandafter\ifx\csname natexlab\endcsname\relax\def\natexlab#1{#1}\fi
\providecommand{\url}[1]{\texttt{#1}}
\providecommand{\href}[2]{#2}
\providecommand{\path}[1]{#1}
\providecommand{\DOIprefix}{doi:}
\providecommand{\ArXivprefix}{arXiv:}
\providecommand{\URLprefix}{URL: }
\providecommand{\Pubmedprefix}{pmid:}
\providecommand{\doi}[1]{\href{http://dx.doi.org/#1}{\path{#1}}}
\providecommand{\Pubmed}[1]{\href{pmid:#1}{\path{#1}}}
\providecommand{\bibinfo}[2]{#2}
\ifx\xfnm\undefined \def\xfnm[#1]{\unskip,\space#1}\fi
\bibitem[{Atsalakis and Valavanis(2009)}]{Atsalakis2009}
\bibinfo{author}{Atsalakis\xfnm[ G.]}, \bibinfo{author}{Valavanis\xfnm[ K.]}.
\newblock \bibinfo{title}{Surveying stock market forecasting techniques –
  part ii: Soft computing methods}.
\newblock \bibinfo{journal}{Expert Systems with Applications}
  \bibinfo{year}{2009};\bibinfo{volume}{36}(\bibinfo{number}{3}):\bibinfo{pages}{5932--5941}.
\newblock \DOIprefix\doi{10.1016/j.eswa.2008.07.006}.
\bibitem[{Atsalakis and Valavanis(2013)}]{Atsalakis2013}
\bibinfo{author}{Atsalakis\xfnm[ G.]}, \bibinfo{author}{Valavanis\xfnm[ K.]}.
\newblock \bibinfo{title}{Surveying stock market forecasting techniques - Part
  I: Conventional methods, In: Zopounidis C, editor. Computation Optimization
  in Economics and Finance Research Compendium}; \bibinfo{publisher}{Nova
  Science, Publishers, Inc, New York}.
\newblock p. \bibinfo{pages}{49--104}.
\bibitem[{Bollerslev(1986)}]{Bollerslev1986}
\bibinfo{author}{Bollerslev\xfnm[ T.]}.
\newblock \bibinfo{title}{Generalized autoregressive conditional
  heteroskedasticity}.
\newblock \bibinfo{journal}{Journal of Econometrics}
  \bibinfo{year}{1986};\bibinfo{volume}{31}(\bibinfo{number}{3}):\bibinfo{pages}{307--327}.
\newblock \DOIprefix\doi{10.1016/0304-4076(86)90063-1}.
\bibitem[{Booth et~al.(2014)Booth, Gerding and Mcgroarty}]{Booth2014}
\bibinfo{author}{Booth\xfnm[ A.]}, \bibinfo{author}{Gerding\xfnm[ E.]},
  \bibinfo{author}{Mcgroarty\xfnm[ F.]}.
\newblock \bibinfo{title}{Automated trading with performance weighted random
  forests and seasonality}.
\newblock \bibinfo{journal}{Expert Systems with Applications}
  \bibinfo{year}{2014};\bibinfo{volume}{41}:\bibinfo{pages}{3651–3661}.
\newblock \DOIprefix\doi{10.1016/j.eswa.2013.12.009}.
\bibitem[{Engle(1982)}]{Engle1982}
\bibinfo{author}{Engle\xfnm[ R.F.]}.
\newblock \bibinfo{title}{Autoregressive conditional heteroscedasticity with
  estimates of the variance of united kingdom inflation}.
\newblock \bibinfo{journal}{Econometrica}
  \bibinfo{year}{1982};\bibinfo{volume}{50}(\bibinfo{number}{4}):\bibinfo{pages}{987--1007}.
\newblock \DOIprefix\doi{10.2307/1912773}.
\bibitem[{Fama(1970)}]{Fama1970}
\bibinfo{author}{Fama\xfnm[ E.F.]}.
\newblock \bibinfo{title}{Efficient capital markets: A review of theory and
  empirical work}.
\newblock \bibinfo{journal}{Journal of Finance}
  \bibinfo{year}{1970};\bibinfo{volume}{25}(\bibinfo{number}{2}):\bibinfo{pages}{383--417}.
\newblock \DOIprefix\doi{110.2307/2325486}.
\bibitem[{Kariya(1993)}]{Kariya1993}
\bibinfo{author}{Kariya\xfnm[ T.]}.
\newblock \bibinfo{title}{Quantitative Methods for Portfolio Analysis}.
\newblock \bibinfo{publisher}{Kluwer Academic Publishers},
  \bibinfo{year}{1993}.
\newblock \DOIprefix\doi{10.1007/978-94-011-1721-0}.
\bibitem[{Nakamula et~al.(2007)Nakamula, Takeo, Okabe and
  Matsuura}]{Nakamula2007}
\bibinfo{author}{Nakamula\xfnm[ S.]}, \bibinfo{author}{Takeo\xfnm[ M.]},
  \bibinfo{author}{Okabe\xfnm[ Y.]}, \bibinfo{author}{Matsuura\xfnm[ M.]}.
\newblock \bibinfo{title}{Automatic seismic wave arrival detection and picking
  with stationary analysis: Application of the km$_2$o-langevin equations}.
\newblock \bibinfo{journal}{Earth, Planets and Space}
  \bibinfo{year}{2007};\bibinfo{volume}{59}(\bibinfo{number}{6}):\bibinfo{pages}{567--577}.
\newblock \DOIprefix\doi{10.1186/BF03352719}.
\bibitem[{Okabe(1999)}]{Okabe1999}
\bibinfo{author}{Okabe\xfnm[ Y.]}.
\newblock \bibinfo{title}{On the theory of km$_2$o-langevin equations for
  stationary flows (1): characterization theorem}.
\newblock \bibinfo{journal}{Journal of the Mathematical Society of Japan}
  \bibinfo{year}{1999};\bibinfo{volume}{51}(\bibinfo{number}{4}):\bibinfo{pages}{817--841}.
\newblock \DOIprefix\doi{10.2969/jmsj/05140817}.
\bibitem[{Okabe(2000)}]{Okabe2000}
\bibinfo{author}{Okabe\xfnm[ Y.]}.
\newblock \bibinfo{title}{On the theory of km$_2$o-langevin equations for
  stationary flows (2): Construction theorem}.
\newblock \bibinfo{journal}{Acta Applicandae Mathematica}
  \bibinfo{year}{2000};\bibinfo{volume}{63}(\bibinfo{number}{1}):\bibinfo{pages}{307--322}.
\newblock \DOIprefix\doi{10.1023/A:1010707332398}.
\bibitem[{Okabe and Inoue(1994)}]{OkabeandInoue1994}
\bibinfo{author}{Okabe\xfnm[ Y.]}, \bibinfo{author}{Inoue\xfnm[ A.]}.
\newblock \bibinfo{title}{The theory of km$_2$o-langevin equations and
  applications to data analysis (ii): Causal analysis (1)}.
\newblock \bibinfo{journal}{Nagoya Mathematical Journal}
  \bibinfo{year}{1994};\bibinfo{volume}{134}:\bibinfo{pages}{1–28}.
\newblock \DOIprefix\doi{10.1017/S0027763000004827}.
\bibitem[{Okabe and Nakano(1991)}]{OkabeandNakano1991}
\bibinfo{author}{Okabe\xfnm[ Y.]}, \bibinfo{author}{Nakano\xfnm[ Y.]}.
\newblock \bibinfo{title}{The theory of km$_2$o-langevin equations and its
  applications to data analysis (i): Stationary analysis}.
\newblock \bibinfo{journal}{Hokkaido Math J}
  \bibinfo{year}{1991};\bibinfo{volume}{20}(\bibinfo{number}{1}):\bibinfo{pages}{45--90}.
\newblock \DOIprefix\doi{10.14492/hokmj/1381413801}.
\bibitem[{Okabe and Yamane(1998)}]{OkabeandYamane1998}
\bibinfo{author}{Okabe\xfnm[ Y.]}, \bibinfo{author}{Yamane\xfnm[ T.]}.
\newblock \bibinfo{title}{The theory of km$_2$o-langevin equations and its
  applications to data analysis (iii): Deterministic analysis}.
\newblock \bibinfo{journal}{Nagoya Mathematical Journal}
  \bibinfo{year}{1998};\bibinfo{volume}{152}:\bibinfo{pages}{175–201}.
\newblock \DOIprefix\doi{10.1017/S002776300000684X}.
\bibitem[{Sakai(2005)}]{Sakai2005}
\bibinfo{author}{Sakai\xfnm[ T.]}.
\newblock \bibinfo{title}{Time series analysis of financial markets : a new
  approach toward econophysics}.
\newblock \bibinfo{publisher}{Doctor's thesis, The University of Tokyo},
  \bibinfo{year}{2005}.
\bibitem[{Suzuki et~al.(2006)Suzuki, Okabe and Fujii}]{Suzuki2006}
\bibinfo{author}{Suzuki\xfnm[ K.]}, \bibinfo{author}{Okabe\xfnm[ Y.]},
  \bibinfo{author}{Fujii\xfnm[ T.]}.
\newblock \bibinfo{title}{On a non-linear risk analysis for stock market
  indexes}.
\newblock \bibinfo{journal}{Asia-Pacific Financial Markets}
  \bibinfo{year}{2006};\bibinfo{volume}{13}(\bibinfo{number}{3}):\bibinfo{pages}{259--259}.
\newblock \DOIprefix\doi{10.1007/s10690-007-9052-y}.

\end{thebibliography}
\end{document}